\documentclass[doublecol]{epl2}

\title{Barkhausen-type noise in the resistance of antiferromagnetic Cr thin films}

\author{L. Tosi\inst{1,2} \and E. Osquiguil\inst{1,2} \and E. E. Kaul\inst{1,2} \and C. A. Balseiro\inst{1,2}}
\shortauthor{L. Tosi \etal}

\institute{
  \inst{1} Centro At\'omico Bariloche, Comisi\'on Nacional de
Energ\'{\i}a At\'omica - 8400 Bariloche, Argentina\\
  \inst{2} Instituto Balseiro, Comisi\'on Nacional de
Energ\'{\i}a At\'omica and Universidad Nacional de Cuyo - 8400 Bariloche, Argentina}


\abstract{We present an experimental study of the changes generated on the electrical resistance $R(T)$ of epitaxial Cr thin films by the transformation of quantized spin density wave domains as the temperature is changed. A characteristic resistance noise appears only within the same temperature region where a cooling-warming cycle in $R(T)$ displays hysteretic behavior. We propose an analysis based on an analogy with the Barkhausen noise seen in ferromagnets. There fluctuations in the magnetization $M(H)$ occur when the magnetic field $H$ is swept. By mapping $M \rightarrow \Psi_0$ and $H \rightarrow T$, where $\Psi_0$ corresponds to the order parameter of the spin density wave, we generalize the Preisach model in terms of a random distribution of {\it resistive hysterons} to explain our results. These hysterons are related to distributions of quantized spin density wave domains with different sizes, local energies and number of nodes.}

\begin{document}

\maketitle

Among simple metals, Cr is the only one that shows an antiferromagnetic
electronic ground state corresponding to an incommensurate spin density
wave (SDW).\cite{overhauser} A single crystal of Cr cooled below the
N\'eel temperature develops domains whose SDW wave vector ${\mathbf Q}$
may be oriented along any of the main crystallographic directions with the
same probability. These domains are separated by walls which can move back and
forth with temperature, or that can be suppressed by cooling in an
applied magnetic field $H$.\cite{fawcett} Using coherent x-ray diffraction
Shpyrko {\it et al.} \cite{shpyrko} measured the noise spectrum at different
temperatures in Cr single crystals produced by domain wall motion
separating domains with different ${\mathbf Q}$ orientations. Independently, Michel {\it et al.}\- \cite{noiseR} observed the presence
of spontaneous fluctuations on the resistance of Cr films, resembling a telegraphic code. The resistance jumps were associated to the thermal motion of two different domain walls: rotations
of ${\mathbf Q}$ domains, and rotation of the polarization vector ${\mathbf \eta}$ within a single ${\mathbf Q}$ domain.
However, it has been shown that in thin films the SDW confinement leads to the quantization of the wave vector ${\mathbf Q}_N$, $N$ being an integer, which orients perpendicular to the film
surfaces.\cite{sonntag,bodeker} This quantization is due to the boundary conditions imposed by the film surfaces or interfaces.
Rich hysteretic phenomena in the electronic transport properties \cite{kummamuru,nosotros} have been
ascribed to the development of domains with different ${\mathbf
Q}_N$ corresponding to a different number of nodes
of the SDW. Consequently, in this case we would expect a different domains structure and dynamics compared to that seen in single crystals or thick films.

In ferromagnetic materials, the existence of domain structures that can move due to the action of an external field $H$ produce the well known Barkhausen noise \cite{barkhausen} in the magnetization $M(H)$. However, in antiferromagnets or SDW systems the noise produced by
domain wall motions as an external parameter is varied, has been more elusive to measure due to the lack of a net magnetic moment.

In this Letter we address an up to now unattended issue related to the resistance changes generated by the
movement or transformation of antiferromagnetic domains that have
different number of nodes (N or N+1) in the quantized SDW state present in
thin Cr films. We show that the whole hysteretic behavior in $R(T)$ is
highly reminiscent of that observed in ferromagnetic materials where
the control field is $H$, and the measured magnitude is the magnetization $M(H)$. In these systems,
Barkhausen noise has been successfully explained in terms of magnetic hysterons
 \cite{hysterons,preisach} using the Preisach model \cite{preisach}. In
 our transport measurements the control field is the temperature $T$ and the measured
 magnitude is the film resistance $R(T)$. Consequently we propose a
 mapping of $H \rightarrow T$ and of $M \rightarrow \Psi_0$, where $\Psi_0$ is
 the order parameter
 associated with the SDW. Using this we show that a simple phenomenological
 extension of the Preisach model based on {\it resistive hysterons} \cite{hysterons}
 reproduce quite satisfactorily the experimental data.
With this model we extract relevant conclusions about the origin of the hysteretic phenomena,
the distribution of energy barriers and the processes involved in
antiferromagnetic domain transformation with temperature. These findings may be relevant to applications in
spintronics, where the understanding of the dynamics of antiferromagnetic domain walls is
of importance to develop better pinning layers, as well as in general problems of antiferromagnets in confined geometries.

\begin{figure*}[t!]
\includegraphics[width=0.98\textwidth]{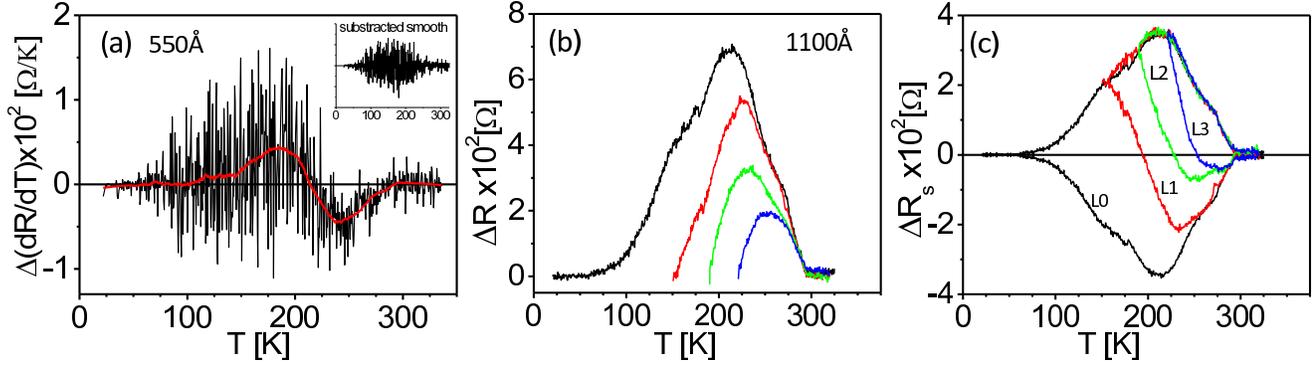}
\caption{(color online) Experimental data: (a) Noise in the difference of derivatives $dR_c/dT - dR_w/dT$
associated with the complete loop in a 550\AA \- thick film. The solid line corresponds
to a smoothing of the data (see text). The inset shows the difference of the data and the smoothing. (b) Difference $\Delta R(T)$ of the resistance in cooling-warming cycles of different amplitudes in $T$ (with final temperature $T_f$), for a 1100\AA \- thick film. (c)
FORCs, L0 ($T_f=$ 50\- K), L1 ($T_f=$ 150\- K), L2 ($T_f=$ 185\- K) and L3 ($T_f=$ 225\- K), calculated from the data in (b) with $\Delta R_s$ as defined in the text.}
\label{experiment}
\end{figure*}

Cr thin films were grown epitaxially on MgO (100) substrates using DC
magnetron sputtering. The films were characterized by x-ray diffraction showing rocking curves
with an angular dispersion at FWHM around the [002] peak of $0.5^\circ$ indicating a very good degree of epitaxy. Results from AFM scans show a mean surface roughness of about 20\AA, which is the third part of the SDW wavelength at low temperatures. The samples were patterned in a four terminal configuration bar of 2\- mm long and 140\- $\mu$m wide using photolithography and chemical
etching. The resistance was
measured using a dc current of 10$\mu$A as a function of $T$ between 20 and 320\- K in steps of 0.5\- K in a commercial cryo-cooler. In each step the temperature was stabilized within 30\- mK giving an intrinsic error
in the resistance measurement between 0.3 to 0.8 m$\Omega$, almost independent of $T$. Each $R(T)$ curve took about 24 hours of measurement.

In Fig.\ref{experiment} we present our main experimental findings. The changes in the resistance produced by the cooling and warming of the sample are small jumps ($|\Delta R| \simeq$ 10$^{-2}$ $\Omega$). Since the resistance is of the order of tens of Ohms ($R$(300\- K) $\simeq$ 63\- $\Omega$) a plot of $R(T)$ shows no appreciable features. Instead, in
panel (a) we show an amplification of the effect by plotting the difference of the resistance derivatives for a complete (20\- K to 320\- K) cooling (\textit{c}) and warming (\textit{w})
  cycle: $dR_c/dT - dR_w/dT$ for a 550\- \AA \- thick film. The
  solid line corresponds to a smoothing of the raw data using the Savistky-Golay method.
  The inset shows the difference between the raw  and smoothed data.
  Although noise is seen in the whole temperature
  range, it is evident that its amplitude strongly increases in the temperature
window where the irreversible behavior in $R(T)$ sets in (see inset). This clearly indicates that
the observed noise in $dR/dT$ is not exclusively due to thermal fluctuations. We claim that it has to be related to the same mechanism
that governs the hysteretic behavior.

  Neutron diffraction \cite{fullerton}, x-ray
  diffraction \cite{kummamuru}, and electrical transport measurements
  \cite{kummamuru,nosotros}, show that the hysteretic behavior in $R(T)$ and $dR(T)/dT$ is due to the
  existence of antiferromagnetic domains with different number of nodes
  $N$ in the confined SDW. These domains switch from $N$ to $N+1$ as the temperature is
  decreased. As we
  will show and elaborate in more detail below, we assume that the
  noise pattern seen in Fig.\ref{experiment} (a) is mainly due to the
  switching and growth of these domains as the temperature of the film
  is changed. From previous results \cite{nosotros} we know that the
  hysteretic region enlarges in $T$ as the film thickness increases.
  Taking this into account, we performed measurements of
  cooling-warming cycles of different amplitudes in $T$ in a Cr film with a thickness
of 1100\AA\- for which the irreversible behavior embraces a wider temperature range as
compared to that found in thinner films.
  The cycles were measured starting always at a temperature of 320\- K,
  cooling down the film to a given final temperature (which we call $T_{f}$),
  and then warming it up again to 320\- K. The results of such measurements plotted as
$\Delta R(T)=R_c - R_w$ are shown in panel (b) of Fig.\ref{experiment}. Clearly, as
$T_{f}$ is increased  the cycles reduce their
   amplitude, the maximum shifts to higher temperatures, and the curves tend to coincide
with the complete cycle
   curve at high enough temperatures. From these data we may
   construct the equivalent to the first order reversal curves (FORCs) \cite{forcM}
in ferromagnets. They are shown in panel (c) and are defined as $\Delta R_s = R_c - R^0_{1/2}$ for the cooling cycle
    and $\Delta R_s = R_w - R^0_{1/2}$ for the warming cycle, where $R^0_{1/2}=(R^0_c +
   R^0_w)/2$ is taken from the complete cycle ($T_{f}=50$\- K).

The measured resistance depends on the domain distribution, the domain wall
structure and the energy barriers in a complex free energy landscape. Due to the problem
complexity we resort to a simple phenomenological model
to describe the physics behind the observed hysteretic behavior. We assume that the building blocks of the model are {\it resistive hysterons}
that correspond to small hysteresis cycles in $R(T)$. These are the equivalent of {\it magnetic hysterons}
commonly used to explain $M(H)$ hysteresis in ferromagnets. There, the Preisach model naturally incorporates the fact that the magnetization increases with magnetic field and uses the magnetic hysterons to account for the increase (decrease) of the total magnetic moment as the field increases (decreases). In our case, as we show below, the resistive hysterons should describe the transition from a high to a low resistance state as the temperature is decreased in order to properly account for the experimental data. From the detailed analysis of Kummamuru's experimental data on electrical transport \cite{kummamuru} it was inferred that the state with N nodes has larger resistance (or a smaller number of effective carriers) than the state with N+1 nodes. Our recent measurements of the Hall coefficient as a function of temperature also support this view showing hysteretic behavior with a higher number of carriers while heating \cite{kaul}. Although this seems counter-intuitive (in the low temperature phase we would expect the amplitude of the SDW to increase and consequently to reduce the number of carriers), numerical simulations (in multi-layer systems with magnetic interfaces) show that the amplitude of the order parameter is drastically reduced when the number of nodes increases by one \cite{fishman}. Based on this evidence, we consider that domains with N+1 nodes have lower resistance than those with N nodes. With our phenomenological description we attempt to account for the fact that the resistance is given by the presence of domain walls and mostly by a contribution from the domains. Hysterons represent a transition from states with N (high $T$) to N+1 (low $T$) nodes. A second source of resistance noise could be the redistribution of domain walls as growth of the domains implies loss of domain-wall scattering. The shape of the resistance hysteresis loop for the complete cooling-warming cycle shows that the dominant effect is the change in the domains´ resistance: $R_c$ is always larger than $R_w$. If the irreversibility were to be dominated by the domain-wall scattering of pinned walls that jump as the domains grow, the shape of the resistance loop would be different and we should expect a temperature interval where $R_w > R_c$.

\begin{figure}[tbp]
\includegraphics[width=0.48\textwidth]{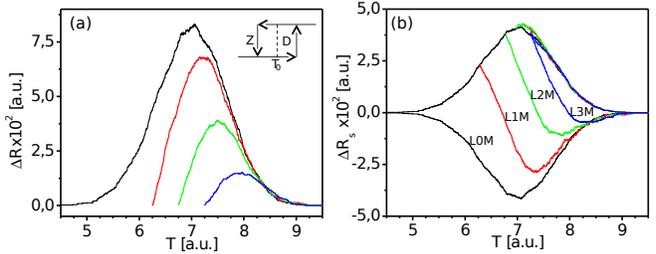}
\caption{(color online) Theoretical model: (a) difference of the resistance in cooling-warming cycles of different amplitudes in $T$. Inset: resistive hysteron. (b) FORC curves L0M, L1M, L2M, and L3M obtained from the data in (a) with $\Delta R_s$ as defined in the text.}
\label{model}
\end{figure}

We characterize these hysterons with three
independent parameters: the center of the loop $T_0$, the half-width of the loop $D$ and the amplitude of
the change in resistance $Z$. See inset in Fig.\ref{model} (a).

Since we are interested in the derivative of $R(T)$ it is better
to mathematically describe the hysterons using an analytical function.
We choose $r(T)= Z \left[
\tanh(\frac{T - T_0 \pm D}{\alpha})+1 \right]$, with $\alpha < D$,
that describes the switching between $2Z$ and $0$ at $T_0\mp D$.
In order to take into account the complex domain structure, we use a collection of independent hysterons with random parameters (${T_0,D,Z}$) each.  Variations stand for different domain sizes, local energies involved and different number of nodes (N or N+1).
Once defined a collection of $N_h$ hysterons, the total resistance at a given temperature $R(T)$ is just the sum of the resistance of each hysteron $r_i(T)$. We can then write

\begin{equation}
R(T)=aT+b+\sum_{i=1}^{N_h} Z_i \left[ \tanh \left( \frac{T - T_{0i} \pm D_i}{\alpha} \right)+1 \right]
\label{eq:resist}
\end{equation}

where the $+$ ($-$) sign stands for cooling (warming), and we have added a linear trend $aT+b$ as the base where hysterons are mounted. This is a
reasonable assumption to account for the phonon and impurity
scattering contributions to the resistance in the temperature range of
interest.

The results obtained from this model are shown in Fig.\ref{model}. Based on
the shape of the FORCs we adopted a random gaussian
distribution for each of the parameters $T_0,D,Z$ with average values and standard deviations: $\langle T_0 \rangle = 7$, $\sigma_{T_0} = 0.6$; $\langle D \rangle = 0.5$, $\sigma_D = 0.12$; and $\langle Z \rangle = 10^{-4}$, $\sigma_Z = 0.15 \times 10^{-4}$. In panel (a) cooling-warming
cycles have been computed using different distributions of hysterons for
each cycle. There are two main effects of $N_h$ on $R(T)$, namely the enhancement of
the difference between $R_c$ and $R_w$ in the hysteresis region, and
the smoothing of the curve shape. When $N_h$ is very small, $N_h\sim
10^1$,  discretization plays a mayor role, giving a steps-shaped curve.
To reproduce the features of our experimental data we took $N_h\sim
700$. In Panel (b) we show the FORCs obtained with the model.

\begin{figure*}[tbp]
\includegraphics[width=0.95\textwidth]{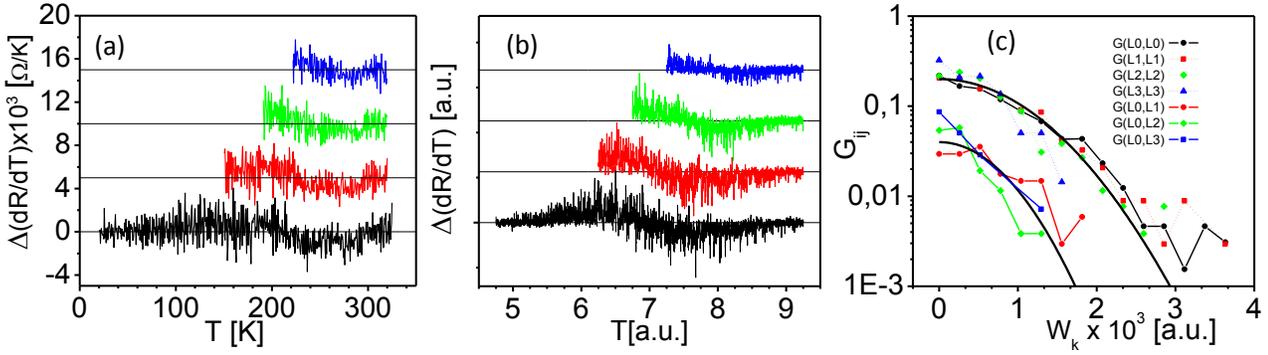}
\caption{(color online) Noise in $d\Delta R/dT$ for the loops shown in: (a) Fig.\ref{experiment} (experiment), and (b) Fig.\ref{model} (model). Panel (c) shows the correlation function $G_{ij}=G\langle Code_i,Code_j\rangle/min\{N_i,N_j\}$ as a function of the window position. See text}.
\label{noiseloops}
\end{figure*}

A comparison of Figs.\ref{experiment} and \ref{model} nicely shows that the results obtained with the model are in very good agreement with the experimental data. This
supports our view of the hysteresis process as a complicated movement of
domain-walls with no memory of the path followed to reach the
state with $N+1$ nodes by lowering the temperature, or with $N$ nodes by
increasing it.

In Fig.\ref{noiseloops} we display the noise in the loops L0 to L3, that is the peak structure present in $d\Delta R/dT$. Panel (a) shows the experimental data shifted vertically by an arbitrary offset for the sake of clarity. In panel (b) the noise patterns computed using the model are plotted. A small random gaussian noise has been included to account for the experimental thermal fluctuations in $R(T)$.
The similarity is remarkable and we may say, using the analogy with
ferromagnets, that this noise in $R(T)$ is the resistive version in
SDW systems of the Barkhausen noise characteristic of ferromagnetic materials.\cite{Kim} However,
we expect some differences between the ferromagnetic and SDW systems. In
particular, the long-range dipolar interaction present in ferromagnetic films
plays an essential role in the domain dynamics. Unfortunately, we do not have
a direct access to observe the dynamics. Since in our case the control
parameter is the temperature, the dynamics of the domain redistribution is affected by the way the sample thermalizes after
each temperature change. Therefore, we only have access to the final state at each temperature.

Figs.\ref{noiseloops} (a) and (b), seem to indicate that each experimental noise pattern is different.

 In order to give a more qualitative analysis of this statement we digitalize the data and introduce a function that allows us to obtain a code-bar type identification of each pattern. We create our code-bar in the temperature axis by dividing it in $M$ small intervals $T_n$. We subtract from each noise pattern in Fig. 3(a) the smoothed curve to obtain a pattern as the one shown in the inset of Fig. 1. Taking the highest (positive or negative) peak amplitude in the resulting pattern we define the relevant scale that is divided in intervals $W_k$ of fixed size $\Delta$. The interval $W_k$ extends from $k\Delta$ to $(k+1)\Delta$ with $k=0,1,...,N$. For the pattern of each loop $L_i$ and each window $W_k$, the code function is defined as:

 \begin{equation}
 \left\{
 \begin{array}{cc} Code_{i,k}[T_n]=1, & if\ the\ peak\ amplitude\ \in W_k\\
  Code_{i,k}[T_n]=0, & otherwise
  \end{array}\right.
 \end{equation}

In this way, for each pattern and window we have an M-dimensional vector whose components are 0 or 1. The square modulus $G^k_{ii}=|Code_{i,k}[Tn]|^2= Code_{i,k}[Tn]\cdot Code_{i,k}[Tn]$ of each code gives the total number of peaks with amplitude in the window $W_k$. In order to compare the noise spectrum of the different loops we first note that although all codes have the same dimension $M$, the first $N_i$ components of loop $L_i$ (i=1,2,3) are zero by definition.  In Fig. \ref{noiseloops}(c) we show the normalized functions $G^k_{ii}/(M-N_i)$  (upper curves) that are equivalent to the histograms of the noise amplitude. The solid line is a fit with a Gaussian function.   In order to analyze the reproducibility of the noise in different loops (Fig.\ref{noiseloops} (a)), we define the correlation  function as the scalar product of two code functions, i.e. $G^k_{ij} = Code_{i,k}[Tn] \cdot Code_{j,k}[Tn]/(M-Nj)$ with $i<j$. The lower curves correspond to $G_{01}$ (full circles), $G_{02}$ (full diamonds) and $G_{03}$ (full squares). The smaller correlation seen in all $G_{0i}$ correspond to stochastic correlations among two different gaussian distributions, one for loop L0, and the others for loops L$_i$, with $i=1,2,3$. Correspondingly, the solid line is a fit with the same upper Gaussian function but squared.

    It is important to realize that if the noise patterns were exactly the same for all loops, irrespective of its temperature span, all curves should coalesce onto a single curve (as clearly is the case for the autocorrelation functions $G_{ii}$). If instead, only part of the loops were correlated (for example the common path when cooling down) the $G^k_{ij}$ should give larger values than the correlations observed among two different gaussian distributions. Therefore, this
    result illustrates in a more quantitative way that each realization of the experiment follows a different path with respect to the distribution of SDW domains, giving rise to different resistive noise patterns in each different run. Finally, we
    mention that the noise amplitude with quasi-Gaussian distribution is present for all studied films, and has a standard deviation that increases with film thickness.

 Another way to analyze in more detail the noise patterns, would be to construct a FORC diagram \cite {forc} with the experimental data. This procedure would allow to estimate the distribution of hysteron parameters directly from the experiments. However, this task would demand more than a year for measuring a reasonable number of loops, say 200, because each loop takes almost 50 hours of measurement. The good agreement between experiments and simulations led us to evaluate the FORC using our simple model. The resulting FORC diagram is similar to the one obtained for systems of single domain ferromagnetic particles (SDFP) with a narrow size distribution. In these systems each particle is subject to the effect of and external magnetic field and to a random field $H_{int}$ due to the dipolar interactions.  In view of these similarities, we can make an analogy between the Cr SDW resistivity hysteretic behavior and the much simpler case of the SDFP.
  In the SDFP system each hysteron corresponds to the orientation of one particle magnetization along the external field direction. The random fields play an important role in determining the shape of the FORC diagram. In the case under study here, hysterons correspond to transitions between different configurations of domains with N or N+1 SDW nodes. The role of the random fields is played by a random distribution of $T_0$ which is crucial in order to reproduce the experimental observations. Finally, the particle size distribution in the SDFP case is here represented by the distribution of resistance jumps $Z$, and the values of $D$ are related to the energy barriers strength between different domain configurations.

In summary, we have shown experimentally that the resistance of thin Cr films displays a characteristic noise in the same temperature region where hysteretic behavior is seen. We presented an extension of the Preisach model and the introduction of resistive hysterons in a simple phenomenological model which reproduced quite satisfactorily the noise patterns in $dR/dT$ and hysteresis in $R(T)$. These are mainly generated by the switching of a random distribution of SDW domains with N or N+1 nodes in the spin density wave. The main assumption supported by previous experiments \cite{kummamuru} is that the domains with N+1 nodes have a lower resistance than those with N nodes. Our results present new evidence supporting this scenario. However, to our knowledge a microscopic theory is still lacking and new theoretical efforts are needed to complete the understanding of this interesting issue. We show that each realization of cooling-warming cycles follows a different path with a different random distribution of domains. We conclude that the system presents a narrow distribution of $T_0$ that suggests a narrow distribution of meta stable domain configurations. Our approach also opens the possibility to study domain evolution as a function of temperature in other antiferromagnetic materials in confined geometries.

\acknowledgments
We thank fruitful discussions with E. Jagla. E.O., E.K. and C.A.B. are members of CONICET. L.T. has a scholarship from CONICET. This work was partially supported by CONICET, by PIP No 11220080101821 of CONICET, by PICT Nos 2006/483 and R1776 of the ANPCyT, Argentina.

\end{document}